\documentclass[iop]{emulateapj}  

\newcommand\aov{\ifmmode{\alpha_{\rm ov}}\else $\alpha_{\rm ov}$\fi}
\newcommand\fov{\ifmmode{f_{\rm ov}}\else $f_{\rm ov}$\fi}
\newcommand\amix{\ifmmode{\alpha_{\rm MLT}}\else $\alpha_{\rm MLT}$\fi}
\newcommand\kms{\ifmmode{\rm km\thinspace s^{-1}}\else km\thinspace s$^{-1}$\fi}
\newcommand\gs{GS98}
\newcommand\as{A09}

\shortauthors{Claret \& Torres}
\shorttitle{Core overshooting}

\begin{document} 

\submitted{Accepted for publication in The Astrophysical Journal}

\title{The dependence of convective core overshooting on stellar mass:
  Additional binary systems and improved calibration}

\author{
Antonio Claret\altaffilmark{1} and
Guillermo Torres\altaffilmark{2}
}

\altaffiltext{1}{Instituto de Astrof\'{\i}sica de Andaluc\'{\i}a,
  CSIC, Apartado 3004, 18080 Granada, Spain}

\altaffiltext{2}{Harvard-Smithsonian Center for Astrophysics, 60
  Garden St., Cambridge, MA 02138, USA}

\begin{abstract}

Many current stellar evolution models assume some dependence of the
strength of convective core overshooting on mass for stars more
massive than 1.1--1.2~$M_{\sun}$, but the adopted shapes for that
relation have remained somewhat arbitrary for lack of strong
observational constraints. In previous work we compared stellar
evolution models to well-measured eclipsing binaries to show that,
when overshooting is implemented as a diffusive process, the fitted
free parameter \fov\ rises sharply up to about 2~$M_{\sun}$, and
remains largely constant thereafter. Here we analyze a new sample of
eight binaries selected to be in the critical mass range below
2~$M_{\sun}$ where \fov\ is changing the most, nearly doubling the
number of individual stars in this regime. This interval is important
because the precise way in which \fov\ changes determines the shape of
isochrones in the turnoff region of $\sim$1--5~Gyr clusters, and can
thus affect their inferred ages.  It also has a significant influence
on estimates of stellar properties for exoplanet hosts, on stellar
population synthesis, and on the detailed modeling of interior stellar
structures, including the calculation of oscillation frequencies that
are observable with asteroseismic techniques.  We find that the
derived \fov\ values for our new sample are consistent with the trend
defined by our earlier determinations, and strengthen the
relation. This provides an opportunity for future series of models to
test the new prescription, grounded on observations, against
independent observations that may constrain overshooting in a
different way.

\end{abstract}


\section{Introduction}
\label{sec:introduction}

Stars more massive than 1.1--1.2~$M_{\sun}$ develop convective cores
and experience the phenomenon of core overshooting, which effectively
enlarges this central region beyond the limits specified by the
classical Schwarzschild criterion. Stellar models computed with
convective core overshooting have longer main-sequence lifetimes, and
have a higher degree of mass concentration toward the center. Such
models have been quite successful at matching the measured physical
properties (masses, radii, temperatures) of double-lined eclipsing
binaries (DLEBs), and have also significantly improved the agreement
between theory and observation regarding the measured rates of apsidal
motion in eccentric DLEBs, compared to historical comparisons
\citep[see, e.g.,][]{Claret:1993, Claret:2010}. Additional examples
illustrating the need for convective core overshooting, mainly in the
context of DLEBs, were given by \citet{CT:16} and references therein.

The last decade or so has seen renewed interest in the subject of
overshooting with a number of efforts directed toward understanding
the underlying physics, discriminating among various proposed
mechanisms, constraining how much overshooting is needed, and
establishing how it might vary as a function of the mass of the star
\citep[see][]{Ribas:2000, Claret:2007, Magic:2010, Aerts:2013,
  Meng:2014, Stancliffe:2015, Valle:2016, Moravveji:2016,
  Deheuvels:2016, Pedersen:2018}. A brief historical overview may be
found in our earlier papers in this series, \cite{CT:16} (Paper~I) and
\cite{CT:17} (Paper~II).

The most recent of these studies used a sample of 29 well-measured
DLEBs to investigate the dependence of core overshooting on stellar
mass in the framework of the increasingly popular diffusive
approximation, with its free parameter \fov\ that controls the extent
of the overshooting layer (\citealt{Freytag:1996, Herwig:1997}; see
also Section~\ref{sec:models} for the mathematical definition of
\fov). It was found that \fov\ increases rapidly in the range between
$\sim$1.2 and $\sim$2.0~$M_{\sun}$ from a value of zero to about
0.016, and changes little thereafter, up to the 4.4~$M_{\sun}$ mass
limit of the observational sample. This result was found to be
independent of the adopted (solar-scaled) element mixture
(\citealt{Grevesse:1998}, or \citealt{Asplund:2009}), even though the
first choice was used in combination with a different primordial
helium abundance and enrichment law than the second. Related to this,
Paper~II also showed that, regardless of the element mixture, the
fractional change in the mass of the convective core at the zero-age
main-sequence (ZAMS) compared to standard (i.e., no-overshooting)
models has a steeper dependence on mass for less massive stars,
leveling off beyond about 2~$M_{\sun}$.  The trend is connected with
the opacities, the equation of state, and the nuclear reaction rates.

The importance of the above study is that it provided the first
reasonably well-established semi-empirical calibration of \fov\ as a
function of stellar mass, at least up to 4.4~$M_{\sun}$. Prior to that
work, most publicly available grids of stellar evolution calculations
used somewhat arbitrary prescriptions to describe the change in the
efficiency of overshooting with mass, or even a constant value for
\fov\ regardless of the mass of the star.

A weakness of the calibration of \fov\ with mass in Paper~II, however,
is the shortage of stars in the critical mass range below about
2~$M_{\sun}$, seen in Figure~2 of that work. This is where
overshooting is changing the most. The way in which \fov\ varies as a
function of mass can have considerable impact on the morphology of
isochrones constructed from evolutionary tracks, and nowhere is this
detailed shape more important than at the turnoff of open clusters.
This region of a cluster's color-magnitude diagram carries most of the
weight for determining its age from classical isochrone fits. Turnoff
masses in the $\sim$1.2--2.0~$M_{\sun}$ interval correspond to cluster
ages roughly in the range 1--5~Gyr, which encompasses a large fraction
of the well-known and best-studied open clusters in the Milky Way. The
impact of overshooting extends to many other fields including
population synthesis, the determination of stellar properties for
exoplanet hosts larger than about 1.2~$M_{\sun}$, and the calculation
of oscillation frequencies that can be measured with asteroseismic
techniques. Indeed, these measurements already exist for large numbers
of stars based on observations from space missions such as {\it
  CoRoT}, {\it Kepler/K2}, and soon others such as {\it TESS} and {\it
  PLATO}.  A first motivation for the present paper is therefore to
enlarge the sample of DLEBs in this critical mass interval in order to
improve the definition of the slope. To this end we examined the
literature and identified seven systems with well-measured properties
that might be added and that are all on the main-sequence, some
reaching near the terminal-age main sequence (TAMS) where sensitivity
to overshooting is greater. We refer to this as our ``main sample''.

Coincidentally, all seven binaries have previously been reported to be
problematic to fit with current models, in the sense that the more
massive primary components were found to be systematically younger
than the secondaries \citep{Clausen:2010, Torres:2014}. In light of
the way in which the strength of overshooting is now known to change
with stellar mass precisely in the mass range of these binaries, our
suspicion is that this may play an important role in those
difficulties. Resolving this lingering issue provides an additional
motivation for our study.

Thirdly, two other classical systems, YZ~Cas and TZ~For, are of
particular interest and may be added as well. YZ~Cas is a moderately
evolved system with components of very different mass (2.26 and
1.32~$M_{\sun}$), one of which is in the range we are most concerned
with here.  The unequal masses provide increased leverage for testing
models, and a unique opportunity to examine the \fov\ vs.\ mass
relation within the same system. Previous studies have concluded that
current models are unable to match all measured properties
simultaneously \citep{Pavlovski:2014}. TZ~For is a rare example of a
binary with one component in the helium-burning phase and the other
less evolved, which has had its absolute masses significantly improved
recently \citep{Gallenne:2016}. Both stars are near 2~$M_{\sun}$,
precisely where the overshooting relation appears to turn over.

Lastly, we take the opportunity to improve our previous fit and
inferred \fov\ values in Paper~II for the evolved eclipsing system
OGLE-LMC-ECL-15260, a pair of giants with indistinguishable masses
around 1.4~$M_{\sun}$ but very different sizes, and to correct a
misprint for OGLE-LMC-ECL-03160 in Table~2 of that work.

The layout of our paper is as follows. Section~\ref{sec:sample}
introduces our new sample of DLEBs. In Section~\ref{sec:models} we
describe the stellar evolution codes employed, and the methodology we
use to infer the values of \fov\ and the mixing length parameter
\amix\ for each star in each system.  The results are then reported
and discussed in Section~\ref{sec:results} based on calculations with
two different element mixtures, and the last section presents our
concluding remarks.

\section{Binary Systems}
\label{sec:sample}

The sample of objects for this study is designed to strengthen the
calibration of \fov\ vs.\ mass in the important but sparsely populated
regime under about 2~$M_{\sun}$, which has the steepest slope.
Initially we selected a total of nine DLEBs not included in Paper~II
that have well measured masses and radii with relative uncertainties
formally smaller than 3\%, as well as measured effective temperatures
and in most cases also spectroscopic metallicities. All are near the
solar abundance, and are evolved enough that their properties are
sufficiently sensitive to the effects of overshooting. Seven of these
systems (V442~Cyg, GX~Gem, BW~Aqr, AQ~Ser, BF~Dra, BK~Peg, and CO~And)
contain main sequence components of similar masses ($q \equiv M_2/M_1
\gtrsim 0.9$) and have been challenging to fit with models in the
past, as mentioned in the Introduction. We are now able to match all
of their properties well except for AQ~Ser, which remains
problematic. We discuss our attempts to fit this system later, but
have chosen to exclude it from our investigation of the dependence of
\fov\ with mass.

The final two systems are TZ~For, in which the primary is a giant, and
YZ~Cas, with components of very different masses ($q = 0.59$) on
either side of the 2~$M_{\sun}$ threshold at which the \fov\ relation
changes slope. The mass determinations for TZ~For have been revised by
\cite{Gallenne:2016} based on new spectroscopic and interferometric
observations since the original study by \cite{Andersen:1991}, and
while the uncertainties are significantly improved, the mass values
are not very different from the previous ones.  \cite{Gallenne:2016}
kept the original radii from \cite{Andersen:1991}, as their study did
not include new light curves. We have made slight adjustments to the
radii here, to account for the fact that the semimajor axis of the
binary changed slightly with the new spectroscopic observations.

\begin{deluxetable*}{lccccc}
\tabletypesize{\scriptsize}
\tablewidth{0pc}
\tablecaption{Binaries systems in our sample.\label{tab:sample}}
\tablehead{
\colhead{Name} & 
\colhead{Mass ($M_{\sun}$)} & 
\colhead{Radius ($R_{\sun}$)} & 
\colhead{$T_{\rm eff}$ (K)} & 
\colhead{[Fe/H]} &
\colhead{Source}
}
\startdata
YZ Cas    & 2.263~$\pm$~0.012    & 2.525~$\pm$~0.011  & 9520~$\pm$~120  &  $+$0.01~$\pm$~0.11  & 1 \\
          & 1.325~$\pm$~0.007    & 1.331~$\pm$~0.006  & 6880~$\pm$~240  &                      &  \\ [1ex]

TZ For    & 2.057~$\pm$~0.001    & 8.34~$\pm$~0.12    & 4930~$\pm$~100  &  $+$0.01~$\pm$~0.04  & 2,3,4 \\
          & 1.958~$\pm$~0.001    & 3.97~$\pm$~0.09    & 6650~$\pm$~200  &                      &  \\ [1ex]

V442 Cyg  & 1.560~$\pm$~0.024    & 2.073~$\pm$~0.034  & 6900~$\pm$~100  &                      & 4,5 \\
          & 1.407~$\pm$~0.023    & 1.663~$\pm$~0.033  & 6800~$\pm$~100  &                      &  \\ [1ex]

GX Gem    & 1.488~$\pm$~0.011    & 2.326~$\pm$~0.012  & 6195~$\pm$~100  &  $-$0.12~$\pm$~0.10  & 4,6 \\
          & 1.467~$\pm$~0.010    & 2.236~$\pm$~0.012  & 6165~$\pm$~100  &                      &  \\ [1ex]

BW Aqr    & 1.479~$\pm$~0.019    & 2.062~$\pm$~0.044  & 6350~$\pm$~100  &  $-$0.07~$\pm$~0.11  & 4,7,8 \\
          & 1.377~$\pm$~0.021    & 1.786~$\pm$~0.043  & 6450~$\pm$~100  &                      &  \\ [1ex]

AQ Ser    & 1.417~$\pm$~0.021    & 2.451~$\pm$~0.027  & 6340~$\pm$~100  &                      & 9 \\
          & 1.346~$\pm$~0.024    & 2.281~$\pm$~0.014  & 6430~$\pm$~100  &                      &  \\ [1ex]

BF Dra    & 1.414~$\pm$~0.003    & 2.086~$\pm$~0.012  & 6360~$\pm$~150  &  $-$0.03~$\pm$~0.15  & 10 \\
          & 1.375~$\pm$~0.003    & 1.922~$\pm$~0.012  & 6400~$\pm$~150  &                      &  \\ [1ex]

BK Peg    & 1.414~$\pm$~0.007    & 1.988~$\pm$~0.008  & 6265~$\pm$~85\phn   &  $-$0.12~$\pm$~0.07  & 8 \\
          & 1.257~$\pm$~0.005    & 1.474~$\pm$~0.017  & 6320~$\pm$~90\phn   &                      &  \\ [1ex]

CO And    & 1.2892~$\pm$~0.0073  & 1.727~$\pm$~0.021  & 6140~$\pm$~130  &  $+$0.01~$\pm$~0.15  & 11  \\
          & 1.2643~$\pm$~0.0073  & 1.694~$\pm$~0.017  & 6170~$\pm$~130  &                      &  
\enddata
\tablecomments{
The first line for each system corresponds to the more evolved
star. In some cases we list \cite{Torres:2010} as an additional
source, as the original determinations were slightly revised in that
work through the use of updated physical constants.  The [Fe/H] value
for YZ~Cas is that of the secondary; the primary is an Am star. The
[Fe/H] value for TZ~For is the weighted average for the primary and
secondary, and the radii have been updated for this work as described
in the text.
Sources are:
(1) \cite{Pavlovski:2014}; 
(2) \cite{Andersen:1991}; 
(3) \cite{Gallenne:2016}; 
(4) \cite{Torres:2010}; 
(5) \cite{Lacy:1987}; 
(6) \cite{Lacy:2008}; 
(7) \cite{Clausen:1991}; 
(8) \cite{Clausen:2010};
(9) \cite{Torres:2014}; 
(10) \cite{Lacy:2012};
(11) \cite{Lacy:2010}.
}
\end{deluxetable*}

The properties of the all systems and the sources for the adopted
values are collected in Table~\ref{tab:sample}, in order of decreasing
primary mass. The eight binaries we retain for this study (i.e., all
except for AQ~Ser) represent a 28\% increase over the sample size
considered in Paper~II. In addition to these eight binaries, we have
revisited OGLE-LMC-ECL-15260 using the same physical properties given
in our earlier study.

\section{Models and Methods}
\label{sec:models}

The principal results for \fov\ reported in this paper are based on
calculations with the Modules for Experiments in Stellar Astrophysics
package \citep[MESA;][]{Paxton:2011, Paxton:2013, Paxton:2015},
version 7385. Microscopic diffusion was included (even for the more
massive of our binaries, including TZ~For) as it is an important
process in the mass range of the present sample. Rotation was not
considered, and mass loss (only relevant for TZ~For) was taken into
account following \cite{Reimers:1977}, with an efficiency coefficient
of $\eta = 0.2$.  Convective core overshooting was treated in the
diffusive approximation, characterized by the free parameter
\fov. Following \cite{Freytag:1996} and \cite{Herwig:1997} the
diffusion coefficient in the overshooting region is given by
\begin{eqnarray}
D_{\rm ov} = D_0\,\exp\left({\frac{-2z}{H_{\nu}}}\right),
\end{eqnarray}
where $D_0$ is the diffusion coefficient at the convective boundary,
$z$ is the geometric distance from the edge of the convective zone,
$H_{\nu}$ is the velocity scale height at the convective boundary
expressed as $H_{\nu} = \fov\ H_p$, and the coefficient \fov\
governs the width of the overshooting layer. The symbol $H_p$
corresponds to the pressure scale height at the edge of the convective
core. The temperature gradient in this region is assumed to be
radiative, and equal values of \fov\ were adopted above the hydrogen-
and helium-burning regions.

For comparison purposes, the analysis was carried out using the two
most common element mixtures for the opacities, by
\cite{Grevesse:1998} (\gs), and \cite{Asplund:2009} (\as). In both
cases we adopted a primordial helium abundance of $Y_p = 0.249$
\citep{Planck:2016} along with a slope for the enrichment law of
$\Delta Y/\Delta Z = 1.67$, unless otherwise indicated.  For stars
with convective envelopes we used the standard mixing-length theory
\citep{Bohm-Vitense:1958} with its usual free parameter \amix.
Although in what follows \amix\ is adjusted independently for each
star, we note for reference that its value calibrated against the Sun
with these models is $\amix = 1.84$ for the \as\ mixture
\citep[see][]{Torres:2015}. The third-degree equation relating the
temperature gradients was solved using the Henyey option in MESA, and
for the condition to set the boundary of the convection zone we
adopted the Schwarzschild criterion.

A second set of models based on the Granada evolutionary code
\citep{Claret:2004} was used both to check for consistency with a
completely independent code, and to perform further tests on some of
the less satisfactory fits with MESA. Diffusion was not considered, as
it is not implemented in the Granada code, and convective core
overshooting follows the classical step-function formulation, rather
than the diffusive approximation we used in MESA. In the classical
prescription the extra distance traveled by convective elements beyond
the boundary of the core is given by $d_{\rm ov} = \aov H_p$, with
\aov\ being a free parameter. In Paper~II we showed that there is a
fairly tight correlation between \aov\ and \fov\ such that $\aov/\fov
\approx 11.36 \pm 0.22$. We use this below as a handy conversion
factor between the two overshooting parameters when comparing results
from the MESA and Granada codes.

For both models the calculations were performed starting from the
pre-main-sequence phase, and in the case of TZ~For with its giant
primary component, they were extended up to the helium-burning phase.
Extensive grids of evolutionary tracks for the measured mass of each
binary component were generated for \fov\ values ranging from 0.000 to
0.025 in steps of 0.002 with MESA, and \aov\ values of 0.00--0.25 with
the Granada code, in steps of 0.02. Mixing length values in both cases
ranged between $\amix = 1.2$ and 2.4, with a resolution of 0.1. For
systems that we found to require it we occasionally extended the
\amix\ dimension up to a value of 2.7. In most cases the chemical
abundance of the binaries has been measured, and the grids adopted
those values. In other cases it was not known, or was uncertain, or
was simply based on fits to stellar evolution models by others, and
additional values were tried.

Our fitting procedure is essentially the same as used in Paper~II, and
the reader is referred to that work for the details. Briefly, the
observational constraints for each star are the masses, radii, and
effective temperatures. A best match to each point in the grid was
sought using a simple $\chi^2$ statistic to infer the optimal values
of \fov\ (or \aov\ for the Granada tracks) and \amix, with the initial
metal abundance $Z$ constrained to be the same for the two stars in
each binary.  After verifying that the preferred matches were similar
between MESA and Granada (accounting for the scaling between \aov\ and
\fov), and that the implied evolutionary states were also the same, we
then computed finer MESA grids tailored to each system and manually
fine-tuned the overshooting and mixing length parameters (often
varying $Z$ as well) to obtain the final best fits.  Throughout this
procedure we required that the ages be the same for the components
within 5\%, to allow for imperfections in the models.  We found
satisfactory fits for all eight of our targets, and the special case
of the AQ~Ser system that we have excluded will be discussed later.
Typical (1$\sigma$) uncertainties for the inferred \fov\ values are
estimated to be 0.004 (0.003 for the giant component of TZ~For, whose
more advanced state makes it more sensitive to overshooting), and 0.20
for \amix. These were determined through experiments in which we
varied one parameter at a time while requiring the predicted radii and
temperatures of the stars to be within their observational
uncertainties, and the ages to be consistent within 5\%, and also by
examining our grids (in which all parameters were varied) in the
vicinity of the best-fit values, again requiring agreement with the
observations.

\section{Results and discussion}
\label{sec:results}

\subsection{V442~Cyg, GX~Gem, BW~Aqr, BF~Dra, BK~Peg,                                                  
and CO~And}

\begin{figure}
\epsscale{1.15}
\plotone{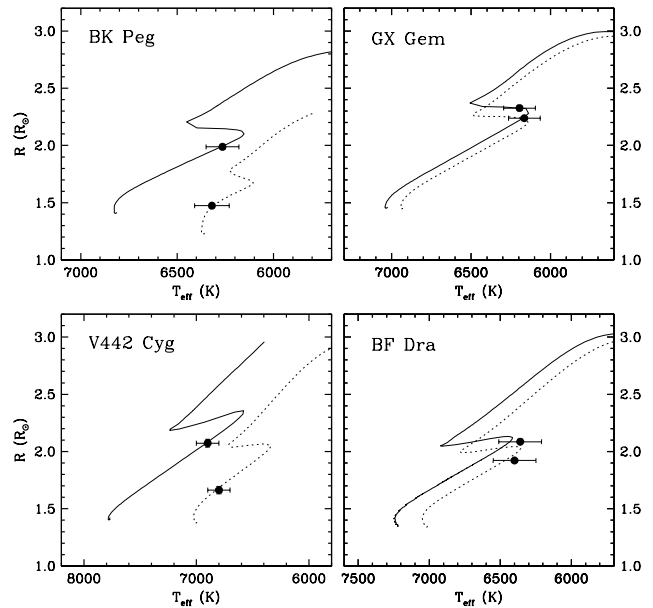}

\figcaption{Four sample fits to evolutionary tracks from the MESA
  models using the \as\ mixture (solid lines for the primary, dotted
  for the secondary). The corresponding model parameters are listed in
  Table~\ref{tab:results}. \label{fig:hr}}

\end{figure}

The inferred values of \fov\ and \amix\ from our best fits are
presented for all our targets in Table~\ref{tab:results}, for both the
\gs\ and \as\ mixtures\footnote{We include OGLE-LMC-ECL-03160 in the
  table, to report a correction to the \gs\ value of \fov\ for the
  primary of the system that was misprinted in Table~2 of Paper~II.}.
There are only minor differences in the derived parameters between the
two sets, which are well within the uncertainties. Also listed are the
best-fit initial abundances, $Z$. Both sets of fitted $Z$ values are
in good agreement with the measured metallicities within their
uncertainties, once due account is taken of the effects of element
diffusion, which tends to decrease the surface abundance through
gravitational settling of heavy elements \citep{Michaud:1970,
  Michaud:1976, Dotter:2017}.  As expected, the abundances derived
with the \gs\ mixture are typically larger than the ones that use \as,
consistent with fact that the corresponding solar abundances are
different.  The mean ages, given in the last column, are
systematically older for \as.  Best fits to four of our targets are
illustrated in Figure~\ref{fig:hr}.

\begin{deluxetable*}{lcccccc}
\tablewidth{0pc}
\tabletypesize{\scriptsize}
\tablecaption{Fitted overshooting and mixing length parameters
using the \gs\ and \as\ mixtures.\label{tab:results}}
\tablehead{
\colhead{} &
\multicolumn{2}{c}{Primary} & 
\multicolumn{2}{c}{Secondary} &
\\
\colhead{Name} &
\colhead{\fov} & 
\colhead{\amix} & 
\colhead{\fov} & 
\colhead{\amix} & 
\colhead{$Z$\tablenotemark{a}} &
\colhead{Mean age (Myr)}
}
\startdata
\multicolumn{7}{c}{\cite{Grevesse:1998} element mixture} \\ [1ex]
YZ Cas                   & 0.016 &  2.00 &  0.006  & 2.74  &  0.012  &    492 \\
YZ Cas\tablenotemark{b}  & 0.0176&  1.66 &  0.0035 & 1.66  &  0.012  &    522 \\
TZ For                   & 0.018 &  1.95 &  0.017  & 2.10  &  0.020  &   1114 \\
V442 Cyg                 & 0.004 &  1.90 &  0.003  & 1.90  &  0.014  &   1409 \\
GX Gem                   & 0.010 &  1.90 &  0.006  & 1.85  &  0.021  &   2349 \\
BW Aqr                   & 0.010 &  2.10 &  0.004  & 1.80  &  0.021  &   2062 \\
BF Dra                   & 0.008 &  1.95 &  0.005  & 1.85  &  0.014  &   2252 \\
BK Peg                   & 0.008 &  1.90 &  0.002  & 2.05  &  0.018  &   2244 \\
CO And                   & 0.003 &  1.93 &  0.000  & 1.72  &  0.016  &   2907 \\
OGLE-LMC-ECL-15260\tablenotemark{c} & 0.004 & 2.03 & 0.004 & 2.11 & 0.006 & 2249 \\
OGLE-LMC-ECL-03160\tablenotemark{d} & 0.008 & 1.94 & 0.008 & 2.15 & 0.0025 & 1023 \\ [1ex]
\multicolumn{7}{c}{\cite{Asplund:2009} element mixture} \\ [+1ex]
YZ Cas                   & 0.015 &  1.70 &  0.005  & 2.67  &  0.010  &    525 \\
YZ Cas\tablenotemark{b}  & 0.0176&  1.66 &  0.0035 & 1.60  &  0.010  &    556 \\
TZ For                   & 0.017 &  1.91 &  0.015  & 1.85  &  0.015  &   1131 \\
V442 Cyg                 & 0.004 &  1.90 &  0.003  & 1.90  &  0.012  &   1482 \\
GX Gem                   & 0.010 &  1.90 &  0.006  & 1.83  &  0.017  &   2397 \\
BW Aqr                   & 0.012 &  1.85 &  0.004  & 1.70  &  0.018  &   2108 \\
BF Dra                   & 0.008 &  1.95 &  0.005  & 1.80  &  0.010  &   2204 \\
BK Peg                   & 0.008 &  1.90 &  0.000  & 2.03  &  0.015  &   2311 \\
CO And                   & 0.002 &  1.90 &  0.000  & 1.72  &  0.014  &   3031 \\
OGLE-LMC-ECL-15260\tablenotemark{c} & 0.004 & 2.08 & 0.004 & 2.08 & 0.004 & 2138
\enddata
\tablecomments{Typical uncertainties are 0.004 for \fov\ (0.003 for
  the giant primary of TZ~For and both components of
  OGLE-LMC-ECL-15260) and 0.20 for \amix.}
\tablenotetext{a}{Bulk (initial) composition.}
\tablenotetext{b}{Preferred parameters, derived using the Granada code
  \citep{Claret:2004} with the step-function approximation for
  overshooting rather than the diffusive approximation, and
  transforming \aov\ to \fov\ using the scale factor $\aov/\fov =
  11.36$ (see text). Values are given to one additional decimal place
  due to the conversion.}
\tablenotetext{c}{Parameters from revised fits that supersede those
  reported in \cite{CT:17}, and place the components in the blue loop
  rather than on the ascending giant branch (see text).}
\tablenotetext{d}{The parameters listed here correct a misprint in Table~2
  by \cite{CT:17} (Paper~II) in the \fov\ value for the primary. Note
  that the helium content for this determination is the one adopted in
  that work, based on $Y_p = 0.24$ and $\Delta Y/\Delta Z = 2.0$,
  rather than the one used here, although this has little influence on \fov.}

\end{deluxetable*}

As indicated earlier, in the past it has not been possible to achieve
entirely satisfactory matches to the physical properties of any of our
seven main targets (V442~Cyg, GX~Gem, BW~Aqr, AQ~Ser, BF~Dra, BK~Peg,
and CO~And) with current models unless the primaries are permitted to
be significantly younger than the secondaries, sometimes by as much as
15\%.  In most cases those fits were carried out with the same amount
of core overshooting for the two components, often a fairly high value
such as $\aov \approx 0.20$ or $\fov \approx 0.020$.  Interestingly,
by now allowing the strength of the overshooting to be different for
each star, we have succeeded in matching all of their properties to
well within the measurement uncertainties, at nearly the same age for
the two components.  The only exception is AQ~Ser, which we discuss
separately below. The masses of all these binaries happen to lie in a
critical regime of the overshooting calibration curve reported in
Paper~II, where \fov\ is not only smaller than mentioned above, but is
also changing rapidly so that even a small difference in mass can
result in a sizable difference in the inferred age of a star when the
behavior of \fov\ is properly taken into account.

\begin{figure}
\epsscale{1.15}
\plotone{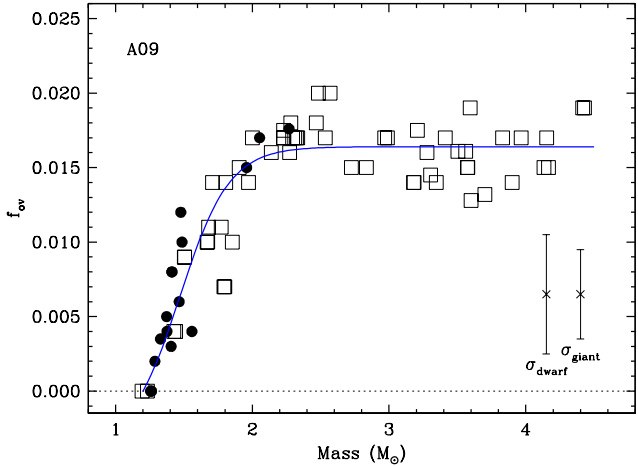}

\figcaption{Inferred \fov\ values from MESA models using the
  \as\ mixture (Table~\ref{tab:results}) as a function of stellar
  mass. Filled circles represent the stars in the present sample
  (including YZ~Cas and TZ~For), and open squares are values taken for
  the 29 dwarf and giant DLEBs from Table~3 of \cite{CT:17},
  determined in the same way with the same element mixture and helium
  content. Typical error bars for dwarfs and giants are indicated on
  the lower right.\label{fig:fov}}

\end{figure}

A graphical representation of our inferred \fov\ values for each star
as a function of stellar mass for the \as\ mixture (filled circles) is
shown in Figure~\ref{fig:fov} see below for a description of the curve
overdrawn), along with the measurements for the 29 other DLEBs
reported by \cite{CT:17} in Paper~II that were derived with the same
element mixture and helium abundance (open squares), and the same
methodology. We include in the figure our determinations for the more
massive binaries YZ~Cas and TZ~For as well (see below). The new
measurements support the general trend found in Paper~II, and are seen
to complement that sample by filling in some of the gaps in the rising
part of the relation. There appears to be one outlier (the primary
component of BW~Aqr) for which we infer a slightly larger \fov\ value
with the \as\ mixture than with the \gs\ one, although the difference
is well within our error bar. This star is also largely responsible
for the hint of a somewhat steeper slope suggested by the current
sample compared to that indicated by the few systems in Paper~II below
2~$M_{\sun}$, though again, we do not consider this hint very
compelling given the uncertainties. Additional, well-measured DLEBs in
the 1--2~$M_{\sun}$ range are needed to investigate this possibility
further.

At the suggestion of the referee we have drawn a curve in
Figure~\ref{fig:fov} that provides a reasonable representation of the
\fov\ measurements. We constrained it by eye to start at
1.2~$M_{\sun}$ and to level off at a value given by the average
\fov\ of all stars with masses above 2~$M_{\sun}$, which is
0.0164. The expression used is
\begin{equation}
\fov = \frac{0.02013}{1+e^{-5.5(M-1.47)}}-0.00373.
\end{equation}
We stress that there is no physical basis for this formula, which is
intended solely to provide a convenient expression for the
overshooting parameter as a function of stellar mass with relatively
few parameters.

\subsubsection{AQ~Ser}

For the AQ~Ser system we were not able to obtain a satisfactory fit to
its measured properties within our 5\% cap for the age difference
between the components, either with MESA or using the Granada code.
With MESA we explored a wide range of chemical compositions (which is
unconstrained observationally) and broader ranges in \fov\ and
\amix\ than for our other binaries, for both element mixtures. We also
performed fits using the temperature \emph{ratio} as an observable
rather than the absolute temperatures of the stars, on the premise
that it is less prone to systematics in DLEB analyses because it does
not depend as strongly on external calibrations, and because it is
more closely tied than the absolute temperatures to the directly
observable difference in eclipse depths. None of these attempts were
successful. We did find an acceptable fit by arbitrarily increasing
the secondary mass; the change required was about twice the
observational error. An additional test with the Granada code involved
using the same metallicity $Z$ for both components, but not forcing
the helium content to be same. This again gave a tolerably good fit,
but at the price of requiring a difference $\Delta Y$ of the order of
10\% between the stars, which would seem unrealistically large.

While this failure of the models could suggest a measurement error in
one or more of the physical properties of the stars ($M$, $R$, or
$T_{\rm eff}$), other causes cannot be ruled out. We note, for
example, that the projected rotational velocity of the primary of
AQ~Ser is quite rapid \citep[$v \sin i = 73 \pm
  10$~\kms;][]{Torres:2014}, while the primaries of both BK~Peg and
BF~Dra have virtually identical masses as the primary of AQ~Ser to
within 0.2\%, but have much slower projected rotational velocities of
$16.6 \pm 0.2$~\kms\ and $10.5 \pm 1.8$~\kms, respectively
\citep{Clausen:2010, Lacy:2012}.  This occurs because of tidal
synchronization, with the orbital period of AQ~Ser being much shorter
(1.69~days) than those of the other two binaries (5.49 and
11.21~days). It is possible that this relatively high rate of rotation
in AQ~Ser~A (shared by the secondary, with $v \sin i = 59 \pm
10$~\kms) may have some effect on the evolution of the system, moving
it away from the canonical evolution the stars would have if they were
single.

\subsection {TZ For and YZ Cas}

TZ~For is a DLEB with well-determined absolute dimensions in an
advanced stage of evolution that makes it useful for this study, and
we have added it to our sample to strengthen the calibration of
\fov\ vs.\ stellar mass. It resembles the much brighter
(non-eclipsing) $\alpha$~Aur binary in that the secondary is in a very
rapid phase of evolution crossing the Hertzprung gap while the primary
is in the helium-burning clump. It represents an interesting example
of significant differential evolution in a system featuring components
of similar mass close to 2~$M_{\sun}$.  \cite{Gallenne:2016} have
recently revisited the mass determinations, which were subsequently
used by \cite{Valle:2017} together with the other binary properties
for a comparison with two sets of evolutionary models, focusing on
inferring the amount of convective core overshooting. One of their
determinations used MESA and the same diffusive overshooting
approximation adopted here with the same \as\ mixture, but with other
minor differences including a fixed solar-calibrated mixing length
parameter of $\amix = 1.74$, and the assumption that the components
have identical \fov\ values. They reported two different solutions,
their preferred one giving an age of $1.10 \pm 0.07$~Gyr and $\fov =
0.013$ with a helium abundance implying $\Delta Y/\Delta Z \approx
1.5$, and the other, poorer fit giving an age of $1.23 \pm 0.03$~Gyr
and $\fov = 0.025$ for $\Delta Y/\Delta Z \approx 1.0$.

\begin{figure}
\epsscale{1.15}
\plotone{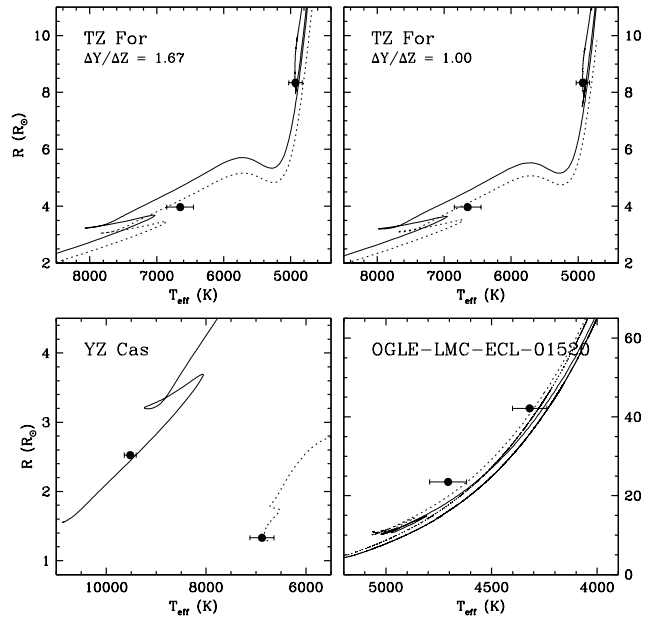}

\figcaption{\emph{Top:} Best-fit solutions with MESA for TZ~For using
  two different enrichment laws ($\Delta Y/\Delta Z$) and the
  \as\ mixture. Primary tracks are drawn with solid lines, secondary
  tracks with dotted lines. Differences are minor. \emph{Bottom:}
  Best-fit solutions for YZ~Cas using the Granada models and the
  step-function approximation to overshooting (\as\ mixture), and for
  OGLE-LMC-ECL-15260 with MESA and the \gs\ mixture, superseding the
  fit in Paper~II that had the stars at a different evolutionary stage
  (see text). For both of these systems $Y_p = 0.249$ and $\Delta
  Y/\Delta Z = 1.67$. \label{fig:hradd}}

\end{figure}

Our own fits to TZ~For with MESA give \fov\ values intermediate
between those above, and are reported in Table~\ref{tab:results} for
the \gs\ and \as\ mixtures and our adopted enrichment law ($Y_p =
0.249$, $\Delta Y/\Delta Z = 1.67$). Our mixing length parameters are
slightly but not significantly super-solar, averaging $\amix \approx
2.0$ for \gs\ and $\amix \approx 1.9$ for \as. To explore the
robustness of these results we performed tests with a different
enrichment law ($\Delta Y/\Delta Z = 1.0$) to match one of the fits by
\cite{Valle:2017}, keeping $Z$ fixed at our best-fit \as\ value. We
obtained essentially the same fits as with our standard enrichment
law, leading also to the same position in the diagram of $R$
vs.\ $T_{\rm eff}$ with the primary in the central helium-burning
phase and the secondary on the subgiant branch (see top panels in
Figure~\ref{fig:hradd}). This suggests \fov\ is rather insensitive to
the helium content, at least in this particular case.

The absolute properties of YZ~Cas, another very interesting DLEB, were
recently redetermined by \cite{Pavlovski:2014} reaching a precision of
0.5\% in the masses and radii of both components. The primary is a
metallic-line A star ($M = 2.263~M_{\sun}$, A2m) that is significantly
more massive than the normal, solar-composition secondary ($M =
1.325~M_{\sun}$, F2).  The extreme mass ratio makes it a uniquely
important object to study the dependence of overshooting as a function
of mass \emph{within the same system}, having one star on the flat
part and the other on the rising part of the \fov\ curve.
\cite{Pavlovski:2014} compared the measured properties of the system
against several models, but reported difficulties finding a
simultaneous match to all of the measurements. Their fits preferred a
significantly lower metallicity than the one they measured for the
secondary (taken to represent the bulk composition of the system, as
the primary is anomalous). Additionally, the ages of the components
were found to be very different (420 and 670~Myr), at least for the
one model for which details were provided.

\begin{figure}
\epsscale{1.15}
\plotone{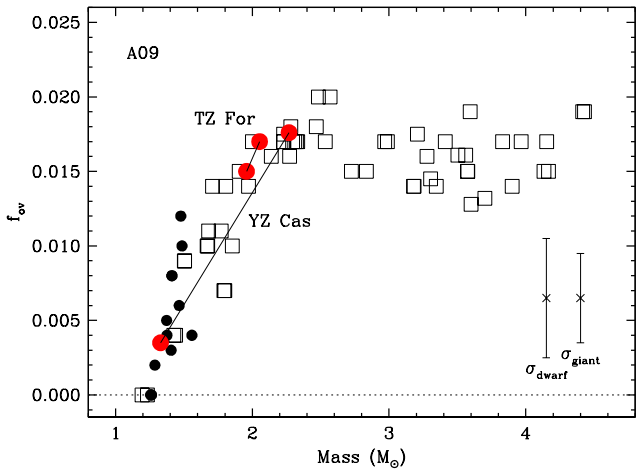}

\figcaption{Same as Figure~\ref{fig:fov} with the \fov\ values for
  TZ~For and YZ~Cas highlighted, and the primary and secondary
  components in each system connected with a line. \label{fig:fov2}}

\end{figure}

Our fits with MESA that allow for independent \fov\ and \amix\ values
for the components still point to a somewhat lower abundance than
measured, but give essentially the same age for the stars, and
\fov\ values that are consistent with the general trend of
Figure~\ref{fig:fov}.  Similar results were obtained with the \gs\ and
\as\ mixtures, which we list in Table~\ref{tab:results}. However, the
mixing length parameter values for the secondary ($\amix \sim 2.7$)
appear implausibly large for a star of this temperature and $\log g$,
and are well outside the range found from the 3-D simulations by
\cite{Magic:2015}. Much more reasonable \amix\ values were found using
the Granada code, which does not account for microscopic diffusion and
treats overshooting in the step-function approximation. The ages are
still very similar, although the lower inferred value of $Z$ persists
with both \gs\ and \as, if somewhat improved. While these fits make
the situation better in some respects, we consider them to be very
provisional pending a better understanding of the remaining
discrepancies. For completeness we list the inferred parameters from
the Granada fits in Table~\ref{tab:results}, where for comparison
purposes we have converted the original \aov\ values (0.20 and 0.04
for the primary and secondary, for both mixtures) to \fov\ values by
means of the scaling constant $\aov/\fov = 11.36$ (see
Section~\ref{sec:models}). These \fov\ values along with those for
TZ~For are highlighted in Figure~\ref{fig:fov2}.

\subsection{OGLE-LMC-ECL-15260 Revisited}

The evolved components in this DLEB have essentially the same mass
($1.426 \pm 0.022$ and $1.440 \pm 0.024~M_{\sun}$) but different radii
of $42.17 \pm 0.33$ and $23.61 \pm 0.69~R_{\sun}$ for the primary and
secondary, respectively \citep{Pietrzynski:2013}. The best fits
presented in Paper~II with the \gs\ and \as\ element mixtures placed
both components on the ascending giant branch, and gave reasonable
matches to their properties. A closer look has revealed that a
somewhat better match can be found with the two components in the blue
loop, a more advanced stage of evolution than we had considered before
(see Figure~\ref{fig:hradd}), which is slower and a priori more
likely.

With the \gs\ mixture the inferred \fov\ and \amix\ parameters are
only slightly altered relative to the ones in Paper~II (and are
consistent within the uncertainties). More importantly, the best-fit
metal content $Z = 0.006$ is now in much better agreement with the
spectroscopically measured metallicity of OGLE-LMC-ECL-15260
(corresponding to $Z = 0.0064$), whereas previously our best fit gave
$Z = 0.003$.\footnote{Note that diffusion has a negligible effect for
  this pair of giant stars.} The change is due in part to our use here
of a different enrichment law than in Paper~II\footnote{In that work
  $Y_p$ and $\Delta Y/\Delta Z$ had been chosen to match our earlier
  study in Paper~I, and enable a proper comparison between \fov\ and
  the \aov\ values of Paper~I.}. With the \as\ mixture we again find a
good fit for identical \fov\ values as in Paper~II, and only a small
increase in the primary \amix\ from 2.00 to 2.08 ($< 1\sigma$). There
is a somewhat more significant decrease in the \amix\ value for the
secondary from 2.25 to 2.08, which brings it into better agreement
with results from 3-D simulations by \cite{Magic:2015}. As before, the
best-fit $Z$ value of 0.004 is consistent with the spectroscopically
measured abundance, which is $Z = 0.0045$ for the \as\ mixture. In
both of these fits the predicted ages of the primary and secondary are
well within 5\% of each other, which meets our requirement
(c.f.\ Section~\ref{sec:models}). The revised parameters are listed in
Table~\ref{tab:results}. For the reasons just described we consider
the new solution for OGLE-LMC-ECL-15260 proposed here, with both
components in the blue loop, to be preferable to the one reported in
Paper~II. The \fov\ values for the \as\ mixture are unchanged, and
those for \gs\ remain perfectly consistent with the general trend of
\fov\ versus mass.

\section{Concluding remarks}
\label{sec:conclusions}

This paper continues our work to investigate the dependence of
convective core overshooting on stellar mass, based on a comparison of
current stellar evolution models with the best available observations
of eclipsing binary systems sufficiently evolved to be useful for this
purpose. Treating overshooting as a diffusive process
\citep{Freytag:1996, Herwig:1997}, we showed in Paper~II that the free
overshooting parameter \fov\ increases rapidly from about
1.2~$M_{\sun}$ to about 2.0~$M_{\sun}$, flattening thereafter. This
result is independent of the element mixture adopted (\gs\ or \as).
Here we have added 8 new binary systems to the sample studied in
Paper~II, designed to improve the coverage in the most important mass
regime below about 2~$M_{\sun}$, where \fov\ is changing the most as
mass increases. The new objects nearly double the number available
earlier in this mass range, adding 15 individual binary components to
the 18 we had before. We have analyzed them with the same methodology
used in our previous work, and they fully support the trend reported
there.

Six of the added systems belong to a group of stars in the narrow mass
range 1.2--1.6~$M_{\sun}$ that had resisted previous attempts to match
their properties at a single age with publicly available stellar
evolution models \citep{Clausen:2010, Torres:2014}. These models
generally offer little or no flexibility to tune the overshooting
parameter, whose dependence on mass is ``hard-wired'' based on various
ad-hoc assumptions or theoretical expectations. By allowing \fov\ to
differ between the components we have succeeded in achieving
satisfactory fits. This supports our suspicion that this was at the
root of the problem, given that all of these stars happen to be on the
steep part of the \fov\ vs.\ mass relation. However, one of our
original targets in this group, AQ~Ser, remains a challenge. We
speculate this may be due to the rapid rotation of the components, or
perhaps measurement errors in some of its properties. A re-examination
of those properties would be helpful.

The YZ~Cas system offers a unique opportunity to check the
\fov\ vs.\ mass relation within the same binary, given that the very
different masses of the components happen to be on either side of the
2~$M_{\sun}$ bending point. We find excellent agreement between the
\fov\ values we infer from our model fits and the trend defined by the
other DLEBs. However, YZ~Cas is not without its problems. Our fits
with the MESA code yield unrealistically high \amix\ values for the
secondary star. The difficulty seems to disappear when using the
Granada models, which do not account for diffusion and have a
different prescription for overshooting. YZ~Cas would benefit from
further study to understand the discrepancies, including perhaps a
check on the measured properties.

Finally, we have revised the solution in Paper~II for
OGLE-LMC-ECL-15260 and found a more satisfactory fit with nearly the
same \fov\ values as before that places both components in the blue
loop rather than on the ascending giant branch, a more likely stage of
evolution.

\begin{acknowledgements}

We are grateful to A.\ Dotter for his assistance in using the MESA
module and for helpful discussions about stellar models. We also thank
the anonymous referee, who provided good suggestions for improving the
manuscript. The Spanish MEC (AYA2015-71718-R and ESP2017-87676-C5-2-R)
is gratefully acknowledged for its support during the development of
this work. GT acknowledges partial support from the NSF through grant
AST-1509375.  This research has made use of the SIMBAD database,
operated at the CDS, Strasbourg, France, and of NASA's Astrophysics
Data System Abstract Service.

\end{acknowledgements}


\begin{thebibliography}

\bibitem[Planck Collaboration(2016)]{Planck:2016} Ade, P.\ A.\ R.,
  Aghanim, N., Arnaud, M.\ et al.\ 2016, \aap, 594, A13

\bibitem[Aerts(2013)]{Aerts:2013} Aerts, C. 2013, in Setting a New
  Standard in the Analysis of Binary Stars, eds.\ K.\ Pavlovski,
  A.\ Tkachenko \& G.\ Torres, EAS Publications Series, Vol.\ 64,
  2013, pp.\ 323-330

\bibitem[Andersen et al.(1991)]{Andersen:1991} Andersen, J., Clausen,
  J.\ V., Nordstr\"om, B., Tomkin, J., \& Mayor, M. 1991, \aap, 246,
  99

\bibitem[Asplund et al.(2009)]{Asplund:2009} Asplund, M., Grevesse, N.,
  Sauval, A.\ J., \& Scott, P. 2009, \araa, 47, 481 (A09)

\bibitem[B\"ohm-Vitense(1958)]{Bohm-Vitense:1958} B\"ohm-Vitense,
  E. 1958, \zap, 46, 108

\bibitem[Claret \& Gim\'enez(1993)]{Claret:1993} Claret, A., Gim\'enez,
A. 1993, \aap, 287, 487

\bibitem[Claret(2004)]{Claret:2004} Claret, A. 2004, \aap, 424, 919

\bibitem[Claret(2007)]{Claret:2007} Claret, A. 2007, \aap, 475, 1019

\bibitem[Claret \& Gim\'enez(2010)]{Claret:2010} Claret, A., Gim\'enez,
A. 2010, \aap, 519, 57

\bibitem[Claret \& Torres(2016)]{CT:16} Claret, A., Torres, G.  2016, 
\aap, 592, A15 (Paper~I)

\bibitem[Claret \& Torres(2017)]{CT:17} Claret, A., Torres, G.  2017, 
\apj, 849, 15 (Paper~II)

\bibitem[Clausen(1991)]{Clausen:1991} Clausen, J.\ V. 1991, \aap, 246,
  397

\bibitem[Clausen et al.(2010)]{Clausen:2010} Clausen, J. V., Frandsen,
  S., Bruntt, H. et al.  2010, \aap, 516, A42

\bibitem[Deheuvels et al.(2016)]{Deheuvels:2016} Deheuvels, S., Brandao, 
I., Silva Aguirre, V., et al. 2016, \aap, 589A, 93D.

\bibitem[Dotter et al.(2017)]{Dotter:2017} Dotter, A., Conroy, C.,
  Cargile, P., \& Asplund, M.\ 2017, \apj, 840, 99

\bibitem[Freytag et al.(1996)]{Freytag:1996} Freytag, B., Ludwig,
  H.-G., \& Steffen, M.\ 1996, \aap, 313, 497
  
\bibitem[Gallenne et al.(2016)]{Gallenne:2016} Gallenne, A.,
  Pietrzy{\'n}ski, G., Graczyk, D.\ et al.\ 2016, \aap, 586, 35

\bibitem[Grevesse \& Sauval(1998)]{Grevesse:1998} Grevesse, N., \&
  Sauval, A.\ J. 1998, Space Sci.\ Rev., 85, 161 (GS98)

\bibitem[Herwig et al.(1997)]{Herwig:1997} Herwig, F., Bloecker, T.,
  Schoenberner, D., \& El Eid, M.\ 1997, \aap, 324, L81

\bibitem[Lacy \& Frueh(1987)]{Lacy:1987} Lacy, C.\ H., \& Frueh,
  M.\ L. 1987, \aj, 94, 712

\bibitem[Lacy et al.(2012)]{Lacy:2012} Lacy, C.\ H.\ S., Torres, G.,
  Fekel, F.\ C., Sabby, J.\ A., \& Claret, A. 2012, \aj, 143, 129

\bibitem[Lacy et al.(2008)]{Lacy:2008} Lacy, C.\ H.\ S., Torres, G.,
  \& Claret, A. 2008, \aj, 135, 1757

\bibitem[Lacy et al.(2010)]{Lacy:2010} Sandberg Lacy, C.\ H., Torres,
  G., Claret, A., Charbonneau, D., O'Donovan, F.\ T., \& Mandushev,
  G. 2010, \aj, 139, 2347

\bibitem[Magic et al.(2010)]{Magic:2010} Magic, Z. Serenelli, A., Weiss, 
A. et al.\ 2010, \apj, 718, 1378  

\bibitem[Magic et al.(2015)]{Magic:2015} Magic, Z., Weiss, A., \&
  Asplund, M.\ 2015, \aap, 573, A89

\bibitem[Meng \& Zhang(2014)]{Meng:2014} Meng, Y., \& Zhang,
  Q.\ S. 2014, \apj, 787, 127

\bibitem[Michaud(1970)]{Michaud:1970} Michaud, G.\ 1970, \apj, 160,
  641

\bibitem[Michaud et al.(1976)]{Michaud:1976} Michaud, G., Charland,
  Y., Vauclair, S., \& Vauclair, G.\ 1976, \apj, 210, 447

\bibitem[Moravveji et al.(2016)]{Moravveji:2016} Moravveji, E.,
  Townsend, R.~H.~D., Aerts, C., \& Mathis, S.\ 2016, \apj, 823, 130

\bibitem[Pavlovski et al.(2014)]{Pavlovski:2014} Pavlovski, K., 
 Southworth, J., Kolbas, V., \& Smalley, B. 2014, \mnras, 438, 590

\bibitem[Paxton et al.(2011)]{Paxton:2011} Paxton, B., Bildsten, L.,
  Dotter, A.\ et al. 2011, \apjs, 192, 3

\bibitem[Paxton et al.(2013)]{Paxton:2013} Paxton, B., Cantiello, M.,
  Arras, P., et al.\ 2013, \apjs, 208, 4

\bibitem[Paxton et al.(2015)]{Paxton:2015} Paxton, B., Marchant, P.,
  Schwab, J., et al.\ 2015, \apjs, 220, 15

\bibitem[Pedersen et al.(2018)]{Pedersen:2018} Pedersen, M.\ G.,
  Aerts, C., P\'apics, P.\ I., \& Rogers, T.\ M. 2018, \aap, in press
  (arXiv:1802.02051)

\bibitem[Pietrzy\'nski et al.(2013)]{Pietrzynski:2013} Pietrzy\'nski,
  G., Graczyk, D., Gieren, W.\ et al. 2013, \nat, 495, 76

\bibitem[Reimers(1977)]{Reimers:1977} Reimers, D. 1977, \aap, 61, 217

\bibitem[Ribas et al.(2000)]{Ribas:2000} Ribas, I., Jordi, C., \&
  Gim\'enez, A. 2000, \mnras, 318, 55

\bibitem[Stancliffe et al.(2015)]{Stancliffe:2015} Stancliffe, R.\ J.,
  Fossati, L., Passy, J.-C., \& Schneider, F.\ R.\ N. 2015, \aap, 575,
  117

\bibitem[Torres et al.(2010)]{Torres:2010} Torres, G., Andersen, J., \&
  Gim\'enez, A. 2010, \aapr, 18, 67

\bibitem[Torres et al.(2015)]{Torres:2015} Torres, G., Claret, A.,
  Pavlovski, K., \& Dotter, A. 2015, \apj, 807, 26

\bibitem[Torres et al.(2014)]{Torres:2014} Torres, G., Vaz, L.\ P.\ R.,
  Lacy, C.\ H.\ S., \& Claret, A. 2014, \aj, 147, 36

\bibitem[Valle et al.(2016)]{Valle:2016} Valle, G., Dell'Omodarme, M.,
  Prada Moroni, P.\ G., \& Degl'Innocenti, S. 2016, \aap, 587, 16

\bibitem[Valle et al.(2017)]{Valle:2017} Valle, G., Dell'Omodarme, M.,
  Prada Moroni, P.~G., \& Degl'Innocenti, S.\ 2017, \aap, 600, A41

\end{thebibliography}
\end{document}